\documentclass[10pt]{article}

\usepackage{amsmath}

\usepackage{array}

\usepackage{appendix}

\usepackage{tocloft}                   

\usepackage{graphicx}

\usepackage{amsfonts}

\usepackage{amssymb}

\usepackage{mathrsfs}

\usepackage{yfonts}

\usepackage{euscript}

\usepackage{centernot}                 

\usepackage{ifsym}                     

\usepackage{upgreek}

\usepackage{mathtools}

\usepackage{color}

\usepackage{slantsc}
\usepackage{calligra}

\usepackage{bbold}          

\usepackage[T1]{fontenc}

\usepackage{epsf}

\usepackage{latexsym}

\usepackage{tipa}

\usepackage{makeidx}

\makeindex

\def\be{\begin{equation}}

\def\ee{\end{equation}}

\def\beq{\begin{equation}}

\def\eeq{\end{equation}}

\def\bea{\begin{eqnarray}}

\def\eea{\end{eqnarray}}

\def\m{\mbox{ }}

\def\mma {\m , \m \m }

\def\!{\hspace{-1.6667em}}

\def\c{\cite}

\def\n{\noindent}

\def\u{\underline}

\def\es{\m = \m}

\def\:={\m := \m}

\def\=:{\m =: \m}

\def\!{\hspace{-1.6667em}}

\def\sbiQ{\mbox{\scriptsize\boldmath$Q$}}

\def\bsigma{\mbox{\boldmath$\sigma$}}                   %

\def\cr{\mbox{\scriptsize{\bf $\m \times \m$}}}

\def\sumi2{\sum\mbox{}_{\mbox{}_{\mbox{\scriptsize $i$=1}}}^2}

\def\sumi3{\sum\mbox{}_{\mbox{}_{\mbox{\scriptsize $i$=1}}}^3}

\def\sumABcycles3{\sum\mbox{}_{\mbox{}_{\mbox{\scriptsize cycles $A,B$=1}}}^{3}}

\def\sumCDcycles3{\sum\mbox{}_{\mbox{}_{\mbox{\scriptsize cycles $C,D$=1}}}^{3}}

\def\sumj3{\sum\mbox{}_{\mbox{}_{\mbox{\scriptsize $j$=1}}}^3}

\def\sumk3{\sum\mbox{}_{\mbox{}_{\mbox{\scriptsize $k$=1}}}^3}

\def\prodiA1{\prod\mbox{}_{\mbox{}_{\mbox{\scriptsize $i$=1}}}^{A - 1}}

\def\bigtimes{\mbox{\Large $\times$}}

\def\pa{\partial}                                                   

\def\FrM{\mbox{\boldmath$\mathfrak{M}$}}                                
\def\lFrg{\mbox{\Large$\mathfrak{g}$}}                         

\def\Hilb{\mbox{{\boldmath$\mathfrak{H}$}ilb}}                 
\def\FrC{\mbox{\Large $\mathfrak{c}$}}                         

\def\scS{\mbox{\scriptsize ${\cal S}$}}                    

\def\FrQ{\mbox{\Large $\mathfrak{q}$}}                               
\def\Phase{\mbox{{\boldmath$\mathfrak{P}$}hase}}                     

\def\bFrR{\mbox{\boldmath$\mathfrak{R}$}}                            
\def\Rig-Phase{\bFrR\mbox{ig-}\Phase}                                
                                                                													   
\def\FrR{\mbox{\boldmath$\mathfrak{R}$}}                             

\def\Positive-Modespace{\mbox{{\boldmath$\mathfrak{M}$}odespace$^+$}}

\def\POSITIVE-MODESPACE{\mbox{{\boldmath$\mathfrak{M}$}ODESPACE$^+$}}

\def\Kin-Hilb{\mbox{{\boldmath$\mathfrak{K}$}in-\Hilb}}                     

\def\Mid-Hilb{\mbox{{\boldmath$\mathfrak{M}$}id-\Hilb}}                     

\def\Dyn-Hilb{\mbox{{\boldmath$\mathfrak{D}$}yn-\Hilb}}                     

\def\5Star{\mbox{\Large$\star$}}              

\begin{document}

\begin{titlepage}

\begin{center}

\LARGE{\bf SHAPE DERIVATIVES} \normalsize

\vspace{0.1in}

\normalsize

\vspace{0.1in}

{\large \bf Edward Anderson$^*$}

\vspace{0.1in}

\end{center}

\begin{abstract}

Shape Theory, together with Shape-and-Scale Theory, comprise Relational Theory.
This consists of $N$-point models on a manifold $\FrM$, 
for which some geometrical automorphism group $\lFrg$ is regarded as meaningless and is thus quotiented out from the $N$-point model's product space $\bigtimes_{I = 1}^N \FrM$.  
Each such model has an associated function space of preserved quantities, solving the PDE system for zero brackets with the sums over $N$ of each of $\lFrg$'s generators.
These are smooth functions of the $N$-point geometrical invariants.
Each $(\FrM, \lFrg)$ pair has moreover a `minimal nontrivially relational unit' value of $N$; 
we now show that relationally-invariant derivatives can be defined on these, yielding the titular notions of shape(-and-scale) derivatives.
We obtain each by Taylor-expanding a functional version of the underlying geometrical invariant, 
and isolating a shape-independent derivative factor in the nontrivial leading-order term.  
We do this for translational, dilational, dilatational and projective geometries in 1-$d$, 
the last of which gives a shape-theoretic rederivation of the Schwarzian derivative.
We next phrase and solve the ODEs for zero and constant values of each derivative.
We then consider translational, dilational, rotational, rotational-and-dilational, Euclidean and equi-top-form (alias unimodular affine) cases in $\geq 2$-$d$.  
We finally pose the PDEs for zero and constant values of each of our $\geq 2$-$d$ derivatives, and solve a subset of these geometrically-motivated PDEs.  
This work is significant for Relational Motion and Background Independence in Theoretical Physics, and foundational for both Flat and Differential Geometry.

\end{abstract}

\n Mathematics keywords: Shape Theory, Foundations of Geometry and Differential Geometry. 

\m 

\n Physics keywords: Background Independence, Configurations and Configuration spaces.  

\m 

\n PACS 04.20.Cv, 04.20.Fy

\m
   
\n $^*$ Dr.E.Anderson.Maths.Physics *at* protonmail.com

\section{Introduction}

\n Shape Theory \cite{Kendall84, Kendall89, Small, Sparr98, Kendall, JM00, M02, MP03, M05, MP05, FORD, GT09, 
                   FileR, Quad, Bhatta, M15, DM16, PE16, KKH16, ABook, ABook2, I, II, III, ACirc, Top-Shapes, Minimal-N, Quad-II, Minimal-N-2}, 
together with Shape-and-Scale Theory \cite{BB82, Iwai87, LR95, LR97, ML00, FileR, A-Monopoles}, 
comprise Relational Theory \cite{FileR, AMech, ABook, ABook, A-Generic}. 
This has the following elements.   

\m 

\n a) A {\it carrier space} manifold 
\be 
\FrC^d := \FrM                                                       \m : 
\ee 
is an incipient model for space's geometrical structure.
In the most commonly considered case, 
\be 
\FrC^d = \mathbb{R}^d                                               \m .
\label{1}
\ee 
Geometry was originally considered to dwell in physical space or objects embedded therein (parchments, the surface of the Earth...).
We consider however the Geometry version of our problem in terms of the abstract carrier space rather than according it an absolute space interpretation.   
In the context of Probability and Statistic, $\FrC^d$ can furthermore be interpreted as a {\it sample space} of {\it location data}.  
In some physical applications, the points model material particles (classical, and taken to be of negligible extent);  
in this context, {\it absolute space} is an alias for carrier space.  
In covering all these settings at once, we refer to points-or-particles. 

\m 

\n b) {\it Constellation space} is the product space 
\be 
\FrQ(\FrC^d, N)  \:=  \bigtimes_{i = 1}^N \FrC^d                    \m ,    
\ee 
where $N$ is the number of points-or-particles under consideration. 
\be 
\mbox{dim}(\FrQ(\FrC^d, N)) = N \, d                                \m , 
\ee 
irrespective of $\FrC^d$. 
For carrier space (\ref{1}), 
\be 
\FrQ(\mathbb{R}^d, N)  \es  \bigtimes_{i = 1}^N \mathbb{R}^d  
                       \es  \mathbb{R}^{N \, d}                     \m .
\ee 

\end{titlepage}

\n c) Some geometrical automorphism group \cite{AMech, Minimal-N, Minimal-N-2} 
\be 
\lFrg = Aut(\FrM, \bsigma)
\ee 
for $\bsigma$ some level of geometrical structure, such as Euclidean, similarity, conformal, affine, projective... 
This is regarded as meaningless and is thus quotiented out from the $N$-point model's constellation space $\FrQ(\FrM, N)$. 
This produces relational space, 
\be 
\FrR(\FrM, N)  \es  \frac{\FrQ(\FrM, N)}{\lFrg}  
               \es  \frac{\bigtimes_{I = 1}^N \FrM}{Aut(\FrM, \bsigma)}   \m . 
\ee 
This is termed more specifically a {\it shape space} if $\bsigma$ includes a scale to be quotiented out, 
                             and a {\it shape-and-scale space} if $\bsigma$ does not, so that scale is retained.  
While some of the most common cases occur in scaled and unscaled pairs, other cases are singleton theories. 
Such singletons have moreover been ascertained to be generic \cite{A-Generic}, due to manifolds generically not admitting a proper similarity Killing vector, 
i.e.\ a similarity Killing vector \cite{Yano70, MacCallum} that is not already a Killing vector. 
Let us finally distinguish between the somewhat earlier \cite{Roach} 1-$d$ clumping purely in terms of length ratios, 
and Kendall's $\geq$ 2-$d$ Shape Theory \cite{Kendall84} which considers relative angle information as well. 

\m

\n Each such model has an associated function space of preserved quantities, 
solving an associated PDE system for zero brackets with the sums over $N$ of each of $\lFrg$'s generators \cite{G63, PE-1, PE-2, PE-3},
\be 
\bf{\mbox [} \scS \bf{\mbox ,} \, \sbiQ \bf{\mbox ]} = 0      \m .  
\ee

\n\be 
\scS  :=  \sum_{I = 1}^N G(\u{q}^I)\frac{\pa}{\pa \u{q}^I}    \m , 
\ee 
Each $(\FrM, \lFrg)$ pair has moreover a `{\it minimal nontrivially relational unit (MRNU)}' \cite{AMech} value of $N$, as explained in Fig 1.  
%
{            \begin{figure}[!ht]
\centering
\includegraphics[width=0.6\textwidth]{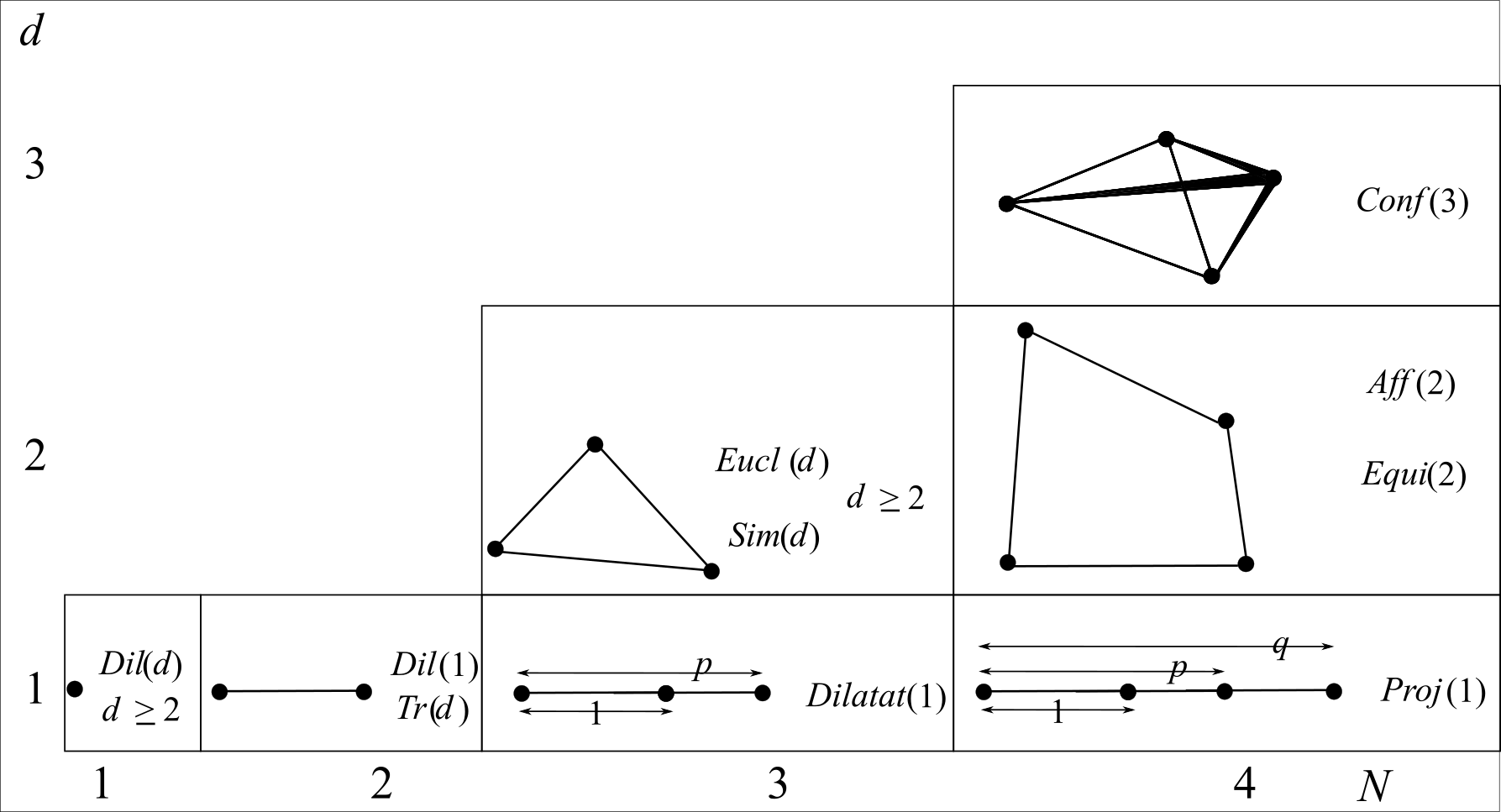}
\caption[Text der im Bilderverzeichnis auftaucht]{        \footnotesize{Grid of smallest $(d, N)$ shape parametrizations, 
indicating which are MNRUs for various flat-space relational theories.}  }
\label{MRNU} \end{figure}          }

\m

\n In the current Article, we show that relationally-invariant derivatives can be defined on these, yielding the titular notions of shape(-and-scale) derivatives.
To this end, we employ a simple method; we first Taylor-expand the functional form of the geometrical invariant functional dependence of the preserved quantities. 
We then isolate a shape-independent derivative factor in the nontrivial leading-order term (Taylor-expanding in enough terms to allow for whichever orders cancel out).  

\m 

\n In Sec 2, we find translational, dilational, dilatational and projective relational derivatives in 1-$d$. 
The last of these amounts to a shape-theoretic rederivation of the Schwarzian derivative \cite{G63, OT05}.  
In Sec 3, we next phrase and solve the ODEs for zero and constant values of each of these derivatives.
We also find the zero derivative cases' solutions to form a progression of geometrically meaningful insights. 

\m 

\n In Sec 4, we find relational derivatives consider the translational, dilational, rotational, rotational-and-dilational and Euclidean cases in $\geq 2$-$d$. 
The unimodular affine transformations, alias equiareal, equivoluminal and equi-top-form transformations in 2-, 3- and arbitrary-$d$, respectively, are also covered.  
In Sec 5, we pose the PDEs for zero and constant values of these derivatives, and solve a subset of these geometrically-motivated PDEs.  
The Conclusion (Sec 6) serves to point out other approaches to (differential) invariants, pointing to various further research projects.  

\m 

\n Let us end by providing some broader motivation for the Relational Theory program that the current Article belongs to.  

\m 

\n{\bf Motivation 1} This Relational Theory and Shape Theory program has Foundations of Geometry \cite{VY10, HC32, S04, Stillwell} applications \cite{PE-1, ABrackets, 9-Pillars}, 
as well as providing answers to geometrical problems \cite{A-Heron, Max-Angle-Flow, Forth}.

\m 

\n{\bf Motivation 2} Kendall's own application was to Shape Statistics, which has to date produced the largest amount of literature 
(see the reviews \cite{Kendall84, Kendall89, Small, Kendall, JM00, Bhatta, DM16, PE16}.

\m 

\n{\bf Motivation 3} The relational side of the Absolute versus Relational Motion Debate \cite{Newton, L, M, BB82, DoD, Buckets} 
is also addressed  \cite{FORD, FileR, Quad, AMech, ABook, ABook2, A-Generic} by this program, 
and there are a number of classical and quantum $N$-Body Problem applications as well \cite{Smale70, Iwai87, LR95, LR97, ML00, M02, M05, FileR, Quad, M15, A-Monopoles, Minimal-N}.

\m 

\n{\bf Motivation 4} It is finally the basis of much recent work \cite{APoT, FileR, APoT2, ABeables, APoT3, AMech, ABook, ABook2, I, II, III, ACirc, Top-Shapes}
on concrete models of Background Independence, which get around many facets of the Problem of Time \cite{DeWitt67, Battelle, A64-A67, Dirac, BB82, HT92, K92, I93, Giu06, FileR, ABook, ABook2} 
and exhibits a strong set \cite{B94I, RWR, ABFKO, FileR, AM13, ABook, ABook2} of analogies with 
the dynamics of General Relativity \cite{ADM, Dirac, DeWitt67, DeWitt70, York74, FM96, Giu09}.

\section{1-$d$ relational derivatives}

\subsection{1-$d$ translationally-invariant geometry's translational derivative}

The invariants in this case are {\it differences}   
\be 
I^{Tr}  :=  x_2 - x_1                                                              \m 
\ee 
(c.f.\ Lagrange and Jacobi coordinates \c{Minimal-N}).
These are based on 2-point MNRU shapes in 1-$d$.  

\m 

\n Taylor-expanding the functional version, 
\be 
{\cal T}r  :=  f(x_0 + \epsilon) - f(x_0)  
            =  \epsilon f^{\prime} + O(\epsilon^2)  \m . 
\ee 
The ensuing {\it translation-invariant derivative} is thus just the {\it ordinary derivative},
\be 
D^{Tr}(f(x)) := f^{\prime}                                                         \m .  
\ee

\subsection{1-$d$ scale-invariant geometry's dilational derivative}

The invariants in this case are {\it ratios}   
\be 
I^{Dil}  :=  \frac{x_2}{x_1}   \m 
\ee
(c.f.\ Euler's inhomogeneity equation of degree zero). 
These are based on 2-point MNRU shapes in 1-$d$. 

\m 

\n Taylor-expanding the functional version,  
\be 
{\cal D}il  \:=  \frac{f(x_0)}{f(x_0 + \epsilon)}  
            \es  1 \m - \m \epsilon \, \frac{f^{\prime}}{f} + O(\epsilon^2) \m . 
\ee 
The {\it scale-invariant derivative} is thus just the {\it logarithmic derivative},
\be 
D^{Dil}(f(x))  \:=  \frac{f^{\prime}}{f}  
               \es  (\mbox{ln} \, f )^{\prime}  \m .  
\ee 
\n{\bf Remark 1} This is homogeneous of degree 1 in its derivatives, and of degree 0 in $f$ itself.

\subsection{1-$d$ dilatational geometry's dilatational derivative}

The invariants are now {\it ratios of differences} \cite{AMech, AObs2}  
\be 
I^{Dilatat}  :=  \frac{x_2 - x_1}{x_3 - x_1}                                                                                  \m .
\ee
These are based on 3-point MNRU shapes in 1-$d$.  

\m 

\n Taylor-expanding the functional version,   
\be 
{\cal D}ilatat(f)  :=  \frac{f(x_0 + \epsilon \, x) -  f(x_0)}{f(x_0 + p \, \epsilon \, x) - f(x_0)}  \es  
\frac{1}{p}\left(   1 \m + \m  \epsilon \, \frac{1 - p}{2}  \frac{  f^{\prime\prime}  }{  f^{\prime}  }  \right) + O(\epsilon^2)  \m . 
\ee
This is valid and nonzero for nondegenerate shapes ($0 \neq p \neq 1$).
The {\it dilational derivative} is thus 
\be 
D^{Dilatat}(f(x))  \:=  \frac{  f^{\prime\prime}  }{  f^{\prime}  }  
         \es  \left(  \mbox{ln} \, f^{\prime}  \right)^{\prime}                                                         \m .  
\label{D-Dilatat}
\ee
\n{\bf Remark 1} This is homogeneous of degree 1 in its derivatives, and of degree 0 in $f$ itself.  

\m 

\n{\bf Remark 2} (\ref{D-Dilatat}) is independent of the choice of $p$, i.e.\ of the 1-$d$ clustering shape formed by the 3 defining points on the line (Fig 1.a).  
It is in this sense that the dilational derivative is shape-independent.  

\m 

\n{\bf Remark 3} This is clearly a composition of the constituent (translational = ordinary) and (dilational = logarithmic) derivatives.  

\m 

\n{\bf Remark 4} (\ref{D-Dilatat}) being second-order, it can be interpreted as a simple notion of curvature \cite{O'Neill, DoCarmo}.

\subsection{1-$d$ Projective Geometry's projective shape derivative}

The invariants are now {\it cross-ratios} \cite{CG67} 
\be
I^{Proj}  :=  [x_1, \, x_2, \, x_3, \, x_4]   \:=   \frac{(w - y)(x - z)}{(w - x)(y - z)}   \m .  
\ee
These are based on 4-point MNRU shapes in 1-$d$. 

\m 

\n Taylor-expanding the functional version,   
\be 
{\cal P}roj  \:=  [f(x_0), \, f(x_0 + \epsilon \, x), \, f(x_0 + p \, \epsilon \, x) , \, f(x_0 + q \, \epsilon \, x)]   
             \es  \frac{p(1 - q)}{p - q} 
\left( 1 - \frac{\epsilon^2}{6} \, q(1 - p) \left(  \frac{f^{\prime\prime\prime}}{f^{\prime}}  - \frac{3}{2}\left( \frac{f^{\prime\prime}}{f^{\prime}} \right)^2  \right)   \right)  
+ O(\epsilon^3)                                                                                                                                                        \m . 
\label{Scw}
\ee 
This is valid and nonzero for nondegenerate shapes ($0$, $p$, $q$, $1$ all distinct).
This gives the {\it projective derivative}, amounting to a rederivation of the {\it Schwarzian derivative} \cite{G63, OT05}
\be 
D^{Proj}(f(x))  \es  {\cal S}(f(x))  
                   \:=  \frac{f^{\prime\prime\prime}}{f^{\prime}}  \m - \m   \frac{3}{2}\left( \frac{f^{\prime\prime}}{f^{\prime}} \right)^2                           \m , 
\ee 
\n{\bf Remark 1} This contains third-order derivatives, and is homogeneous of degree zero in $f$ and of degree 2 in its derivatives.

\m 

\n{\bf Remark 2} (\ref{Scw}) is independent of the choice of $p$ or $q$, i.e.\ of the 1-$d$ clustering shape that the 4 defining points on the line form (Fig 1.b). 

\m 

\n{\bf Remark 3} Some useful alternative expressions for (\ref{Scw}) are as follows. 
\be 
D^{Proj}(f(x))  \es  \frac{    2 \, f^{\prime\prime\prime} f^{\prime} - 3 \left(  f^{\prime\prime}  \right)^2    }{    2 \, f^{\prime \, 2}    } 
         \es  \left(    \frac{  f^{\prime\prime}  }{  f^{\prime}  }  \right)^{\prime} \m - \m  \frac{1}{2} \left(  \frac{  f^{\prime\prime}  }{  f^{\prime}  }  \right)^2 
         \es  \left(    \frac{Y^{\prime}}{Y}     \right)^{\prime}  \m - \m \frac{1}{2}   \left(   \frac{Y^{\prime}}{Y}   \right)^2
		 \es  F^{\prime} - \frac{F^2}{2}                                                                                                                               \m , 
\ee 
for changes of variables 
\be 
Y := f^{\prime}
\label{Y-sub}
\ee  
and 
\be 
F  \:=  \frac{  f^{\prime\prime}  }{  f^{\prime}  } 
   \es  \left(\mbox{ln} \, f^{\prime}\right)^{\prime}                                                                                                                  \m . 
\label{F-sub}
\ee

\section{Corresponding ODEs for zero and constant derivatives}

\subsection{Translational derivatives}

Firstly, zero difference derivative gives the ODE 
\be 
0 = D^{Tr}(f) 
  = f^{\prime}    \m .  
\ee  
This is solved by 
\be 
f = const \m : \m \mbox{ constant functions } . 
\ee
Secondly, constant difference derivative yields the ODE 
\be 
c = D^{Tr}(f) 
  = f^{\prime}     \m . 
\ee 
This is solved by 
\be 
f = c \, x + d \m : \m \mbox{ linear functions } .  
\ee

\subsection{Dilational derivatives}

\n Firstly, zero ratio derivative gives the ODE 
\be 
0  =   D^{Dil}(f)  
  \es  \frac{f^{\prime}}{f}                                  \m . 
\ee 
For 
\be 
f \neq 0                                                     \m , 
\ee 
this collapses to just 
\be 
f^{\prime} = 0                                               \m , 
\ee 
so it is solved by nonzero constant functions.  

\m 
 
\n Secondly, constant ratio derivative yields the ODE 
\be 
c   =   D^{Dil}(f)  
   \es  \frac{f^{\prime}}{f} 
   \es  (\mbox{ln} \, f )^{\prime}                           \m , 
\ee
Integrating once,  
\be 
\mbox{ln} \, f = c \, x + d 
\ee 
Thus 
\be 
f = D \, \mbox{exp} (C \, x)  
\ee 
solves: exponential linear functions.

\subsection{Dilatational derivatives}

\n Firstly, zero dilational derivative yields the ODE 
\be 
0   =  D^{Dilatat}(f) 
   \es  \frac{f^{\prime\prime}}{f^{\prime}}                                                    \m .  
\ee 
For 
\be
f^{\prime} \neq 0                                                                              \m , 
\ee 
this collapses to just 
\be 
f^{\prime\prime} = 0                                                                           \m .  
\ee 
Thus it is solved by nonzero-derivative linear functions, 
\be 
f = c \, x + d \mma  c \neq 0  \m .  
\ee 
This moreover receives the interpretation that linear functions are straight, i.e.\ have zero curvature.  

\m 
 
\n Secondly, constant dilational derivative yields the ODE 
\be 
c   =   D^{Dilatat}(f)  
   \es  \frac{f^{\prime\prime}}{f^{\prime}} 
   \es (\mbox{ln} \, f^{\prime} )^{\prime}                                                     \m .  
\ee
Integrating once,  
\be 
\mbox{ln} \, f^{\prime} = c \, x + d                                                           \m . 
\ee 
Thus 
\be
f^{\prime} = D \, \mbox{exp} (C \, x)                                                          \m , 
\ee 
so integrating again, 
\be 
f = A \, \mbox{exp}(C \, x) + B                                                                \m .  
\ee

\subsection{Projective shape derivatives}

\n Zero projective shape derivative alias Schwarzian derivative gives the ODE 
\be 
0   =  D^{Proj} (f) 
   \es  \frac{  f^{\prime\prime\prime}  }{  f^{\prime}  }  \m - \m \frac{3}{2}\left(\frac{  f^{\prime\prime}  }{  f^{\prime}  }\right)^2                                  \m , 
\ee
i.e.\ 
\be 
2 \, f^{\prime\prime\prime} f^{\prime}  \es  3 \, f^{\prime\prime \, 2}                                                                                                   \m :  
\ee 
a homogeneous-quadratic third-order ODE. 
This is well-known to be solved precisely by the fractional-linear functions
\be 
f  \es  \frac{M \, x + N}{O \, x + P}        \m , 
\label{F-L}
\ee 
We ascertain this by making the substitutions (\ref{Y-sub}, \ref{F-sub}).
This leaves us with 
\be  
F^{\prime}  \es  \frac{F^2}{2}                           \m ,
\ee
so 
\be
x  \es  A - \frac{2}{F}                                  \m , 
\ee
thus inverting, 
\be
\frac{2}{A - x}  \es  F  
                 \es  (\mbox{ln} \, Y)^{\prime}          \m . 
\ee
A second integration gives  
\be 
\mbox{ln} \, Y = -2 \, \mbox{ln}(A - x) + b              \m .
\ee
Thus 
\be 
\frac{B}{(A - x)^2} = Y = f^{\prime}                     \m . 
\ee
A third and final integration then gives  
\be
f  \es  -\frac{B}{A - x} + C                             \m , 
\ee
which, placing under a common denominator, recovers the fractional-linear form (\ref{F-L}).  
This result can be interpreted as the fractional-linear functions playing an analogous role in the Projective Geometry of curves to that of  
                                  the linear functions in the ordinary geometry of curves. 
I.e.\ of projectively-flat curves displaying none of the associated notion of projective curvature.  

\m 
 
\n On the other hand, constant projective shape derivative yields the ODE 
\be 
c    =   D^{Proj}(f)  
    \es  \frac{f^{\prime\prime}}{f^{\prime}}  \m - \m  \frac{3}{2}\left(\frac{f^{\prime\prime}}{f^{\prime}}\right)^2                \m , 
\ee 
i.e.\ the homogeneous-quadratic third-order ODE 
\be 
2 \, f^{\prime\prime\prime} f^{\prime} - 3 \, f^{\prime\prime \, 2} \m = \m  c \, f^{\prime \, 2}                                   \m .  
\ee 
The above pair of substitutions continues to work, yielding 
\be 
x  \es  \sqrt{\frac{2}{c}} \,  \mbox{arctan} \left( \frac{F}{\sqrt{2 \, c}} \right) \m + \m  a                                      \m . 
\ee 
Thus 
\be 
(\mbox{ln}\,Y)^{\prime}   =      F 
                         \es  \sqrt{2 \, c} \, \mbox{tan}\left(\sqrt{\frac{c}{2}}\,  x + A \right)                                  \m , 
\ee 
so a second integration gives  
\be 
\mbox{ln} \, Y  \es  \mbox{ln} \left(  B \, \mbox{sec}^2\left(\sqrt{\frac{c}{2}}\,  x + A \right)  \right)                          \m . 
\ee 
Finally exponentiating both sides and performing a third integration,  
\be 
f \es  B \, \mbox{tan}\left(\sqrt{\frac{c}{2}}\,  x + A \right)  + C                                                                \m . 
\ee

\section{Higher-$d$ examples}

We consider this so as to have relative angle information alongside length ratio information, rather than just the latter, 
by which more than just the mathematics of clumping is required to describe (scaled) shapes.

\subsection{$N$-$d$ translation-invariant geometry's translational derivative}

The invariants in this case are {\it vectorial differences} 
\be
I^{Tr}  \:=  \u{x}_1 - \u{x}_0 
        \:= \u{y}_1  \m .
\ee
These are based on 2-point MNRU shapes.

\m 

\n Taylor-expanding the functional version of this for 
\be 
\u{x}_1 := \u{x}_0 + \epsilon \, \u{y}_1 \m , 
\ee 
\be
{\cal T}r  \:=  \u{f}(\u{x}_1) - \u{f}(\u{x}_0) 
           \es  \epsilon \u{y}_1^a \cdot \pa_a \u{f} + O(\epsilon) \m  
\ee
gives the translational derivative to be 
\be 
D^{Tr}_a \u{f} = \pa_a \u{f}                                     \m :
\ee
the gradient operator acting on a vector.

\subsection{$N$-$d$ scale-invariant geometry's dilational derivative}

The invariants in this case are {\it ratios of components}  
\be
I^{Dil}  \:=  \frac{x_1^a}{x_0^b}  \m .
\ee
These are based on 2-point MNRU shapes (and more occasionally on a single point: when $x_0 = x_1$ but $a \neq b$).

\m 

\n Taylor-expanding the functional version of this, 
\be
{\cal D}il  \es  \frac{f_a(\u{x}_1)}{f_b(\u{x}_0)}  
            \es  1  \m + \m  \epsilon \, \u{y}_1 \frac{\pa_c f_b}{f_a}  \m + \m  O(\epsilon^2)                                  \m .  
\ee
The dilational derivative is thus 
\be 
{D^{Dil \, a}}_{bc}(\u{f}) = \frac{\pa_c f_b}{f_a}                                                         \m , 
\ee
This remains homogeneous of degree zero, but is no longer in general of logarithmic form [c.f.\ (\ref{D-Dilatat})] due to the fixed $a$ and $b$ indices taking distinct values.

\subsection{$N$-$d$ rotation-invariant geometry's rotational derivative}

The invariants in this case are {\it dot products}  
\be
I^{Rot}  \:=  (\u{x} \cdot \u{y}) = \delta_{ab} x^a y^b                                                                                                              \m . 
\ee
These are based on 2-point MNRU shapes.

\m 

\n Taylor-expanding the functional version of this, 
\be
{\cal R}ot  \:=   \u{f}(\u{x}_0) \cdot \u{f}(\u{x}_1)  
            \es  ||\u{f}||^2 + \epsilon \, \u{f} \cdot \u{x}_1^a \pa_a \u{f} + O(\epsilon)             \m . 
\ee
The rotational derivative is thus 
\be 
D^{Rot}_a (\u{f})  \es  \u{f} \cdot \pa_a \u{f}  
                   \es  \pa_a\left(\frac{||\u{f}||^2}{2}\right)                                    \m .  
\ee

\subsection{$N$-$d$ $Rot \times Dil$-invariant geometry}

The invariants in this case are {\it ratios of dot products}
\be
I^{Rot \times Dil}  \:=  \frac{(\u{w} \cdot \u{x})}{(\u{y} \cdot \u{z})}  \es  \frac{\delta_{ab} x^a y^b}{\delta_{cd} x^c y^d}                                                                \m .
\ee
These are based on 4, 3, 2 or even 1 point, depending on how many of the components involved belong to the same vector.  

\m 

\n Taylor-expanding the functional version of this,
\be
{\cal R}ot\mbox{-}{\cal D}il  \:=  \frac{\u{f}(\u{x}_0) \cdot \u{f}(\u{x}_1)}{\u{f}(\u{x}_0) \cdot \u{f}(\u{x}_p)}
                              \es  1 + \frac{\epsilon}{2||\u{f}||^2}(\u{x}_1 - \u{x}_p) \cdot \u{\nabla} ||\u{f}||^2                                                 \m .
\ee 
The rotational-and-dilational derivative is thus 
\be
D_a^{Rot \times Dil}(\u{f}) = \pa_a (\mbox{ln} ||\u{f} ||^2)                                                                                                                \m .
\ee

\subsection{$N$-$d$ Euclidean geometry's relational derivative}

The invariants are now {\it dot products of differences}, 
\be 
I^{Eucl}  \:=  ((\u{w} - \u{y})\cdot (\u{x} - \u{y}) ) = \delta_{ab} (w^a - y^a)(x^b - y^b)                                                                                                    \m . 
\ee
These are based on 3-point MNRU shapes, i.e.\ the triangular MNRU. 
  
\m 
  
\n Taylor-expanding the functional version,   
\be 
{\cal E}ucl  \:=  \delta_{ab}(f^a(x_1) - f^a(x_0))(f^b(x_p) - f_b(x_0))  
          \es  \epsilon \, y_1^a y_p^b (\pa_a \u{f} \cdot \pa_b \u{f})  + O(\epsilon^3)                                                                              \m . 
\ee 
The Euclidean relational derivative is then 
\be 
D^{Eucl}_{ij}(\u{f})  \:=  (\pa_i \u{f} \cdot \pa_j \u{f})  
                      \es   h_{im} {h_j}^m                                                                                                                           \m , 
\ee 
for 
\be 
h_{ik}  := \pa_i f_k                                                                                                                                                 \m .
\ee

\subsection{2-$d$ Equiareal Geometry's relational derivative}

The invariants in this case \cite{Coxeter} are {\it areas formed from differences of vectors},  
\be 
I^{Equi(2)}  \:=  ( (\u{w} - \u{y}) \cr (\u{x} - \u{y} ) )_{\perp}                                               \m . 
\ee 
These are based on between 3 and 4 distinct points, 3 being the triangular MNRU.

\m

\n Taylor-expanding the functional version, 
\be 
{\cal E}qui(2)  :=  \epsilon_{3ab}(f^a(x_1) - f^a(x_0))(f^b(x_p) - f_b(x_0))  
                 =  \epsilon^2 y_1^a y_p^b \left( \pa_a \u{f} \cr \pa_b \u{f} \right)_{\perp}  + O(\epsilon^3)   \m . 
\ee 
The equiareal relational derivative is then 
\be 
D^{Equi(2)}_{ab}(\u{f})  \:=  \left( \pa_a \u{f} \cr \pa_b \u{f} \right)_{\perp}                                 \m . 
\ee

\subsection{3-$d$ Equivoluminal Geometry's relational derivative}

The invariants in this case are {\it volumes -- scalar triple products -- of differences of vectors}, 
\be 
I^{Equi(3)}  \:=  (\u{w} - \u{z}, \, \u{x} - \u{z} , \, \u{y} - \u{z})                                              \m . 
\ee 
These are based on between 4 and 6 distinct points, 4 being the tetrahaedral MNRU.

\m 

\n Taylor-expanding the functional version, 
$$ 
{\cal E}qui(3)   \:=  \epsilon_{abc}(f^a(x_1) - f^a(x_0))(f^b(x_p) - f_b(x_0)) (f^c(x_q) - f_c(x_0))  
$$
\be
          \es  \epsilon^3   y_1^a y_p^b y_q^c (  \pa_a f , \, \pa_b f  , \,  \pa_c  f )           \m . 
\ee 
The equivoluminal relational derivative is then 
\be 
D^{Equi(3)}_{ijk}(\u{f})  \:=  \left(  \pa_a \u{f} , \, \pa_b \u{f} , \, \pa_c \u{f} \right)       \m . 
\ee

\subsection{Arbitrary-$d$ Equi-Top-Voluminal Geometry's relational derivative}

The invariants in this case are {\it top forms of differences} \cite{AMech},\footnote{`Top form' refers to top form supported in dimension $d$, i.e.\ $d$-forms.  
These include area in 2-$d$ and volume in 3-$d$, thus recovering the previous two subsection.} 
\be 
I^{Equi(d)}  \:=   \bigwedge_{a = 1}^d (\u{x}^a - \u{z})                                                                    \m . 
\ee
These are based on between $d + 1$ and 2-$d$ distinct points, $d + 1$ supporting the $d$-simplex of relative vectors MNRU.

\m 

\n Taylor-expanding the functional version, 
$$ 
{\cal E}qui(d)  \:=  \epsilon_{a_1 \m . \m . \m . \m  a_d}(f^{a_1}(x_1) - f^{a_0}(x_0)) \m . \m . \m . \m  (f^{a_d}(x_q) - f^{a_d}(x_0))  
$$
\be
            \es   \epsilon^d       y_1^{a_1} \m . \m . \m . \m  y_{p_{d - 1}}^{a_d} \bigwedge_{i = 1}^d  \pa_{a_i} f       \m . 
\ee 
The equi-top-form relational derivative is then 
\be 
D^{Equi(d)}_{i_1 \m . \m . \m . \m  i_d}  \es  \bigwedge_{i = 1}^d \pa_{a_i}  f                                            \m .  
\ee

\section{Corresponding PDEs for zero and constant versions}

\subsection{Translational derivatives}

Firstly, zero translational derivative yields the PDE 
\be 
0 = D^{Tr}_a \u{f} = \pa_a  (\u{f})       \m .  
\ee
This is solved by  
\be 
\u{f} = \u{c}                           \m : \m \mbox{ constant vector functions} \m . 
\ee
Secondly, constant translational derivative yields the PDE 
\be 
\u{c}_a  =  D^{Tr}_a \u{f} = \pa_a  (\u{f})   \m .  
\ee
This is solved by  
\be 
\u{f}    =  \u{c}_a x^a + \u{d}              \m :   \m \mbox{ linear vector functions} \m . 
\ee

\subsection{Dilational derivatives}

Firstly, zero dilational derivative yields the PDE 
\be 
0  \es  {D^{Dil \, a}}_{bc} (\u{f})  
   \es  \frac{\pa_c f_b}{f_a}                                            \m , 
\ee
so 
\be 
\u{f} = \u{c}                                                            \m :  \m \mbox{ constant vector functions} \m . 
\ee 
\n Secondly, constant dilational derivative yields the PDE 
\be 
{c^a}_{bc} f_a = \pa_c f_b  \m \mbox{ (no sum }                      \m , 
\ee 
i.e.\ 
\be 
\left(  \sum_{a = 1}^N{\delta^a}_b\pa_c  - {c^a}_{bc}  \right) f_a   =  0                         \m .
\ee 
This is a homogeneous linear, overdetermined system, so we need to look for integrable subcases.

\subsection{Rotational derivatives}

Firstly, zero rotational derivative yields the PDE 
\be 
0   =   D^{Rot}_a (\u{f}) 
   \es  \pa_a \left(  \frac{||\u{f}||^2}{2}  \right)    \m , 
\ee
This is solved by 
\be 
||\u{f}|| = const                                       \m : \m   \mbox{ constant-norm vector functions}  \m . 
\ee
\n Secondly, constant rotational derivative yields the PDE 
\be 
c_1  =   D^{Rot}_a  (\u{f})   
    \es  \pa_a \left(  \frac{||\u{f}||^2}{2}  \right)   \m , 
\ee
This is solved by 
\be 
||\u{f}|| = \sqrt{ c_a x^a + d}                           \m : \m \mbox{  vector functions with linear squared-norm} \m .
\ee

\subsection{rotational-and-dilational shape derivatives}

Firstly, zero $Rot \times Dil$ derivative yields the PDE 
\be 
0   =   D^{Rot \times Dil}_a (\u{f}) 
   \es  \pa_a \left(  \mbox{ln} ||\u{f}||^2  \right)    \m .  
\ee
This is solved by  
\be 
||\u{f}|| = const                                                                         \m :  \m \mbox{ constant-norm vector functions } \m . 
\ee 
\n Secondly, constant $Rot \times Dil$ derivative yields the PDE 
\be 
c_a  =  D^{Rot \times Dil}_a (\u{f})  
     =  \pa_a \left(  \mbox{ln} ||\u{f}||^2  \right)                                       \m .  
\ee
This is solved by 
\be 
||\u{f}|| = D \, \mbox{exp}  \left( \frac{c_a x^a}{2}   \right)                    \m :  \m`\mbox{  vector functions of exponential-linear norm }  \m . 
\ee

\subsection{Euclidean relational derivatives}

Firstly, zero Euclidean derivative yields the PDE 
\be 
0  =   D^{Eucl}_{ij}(\u{f})  
  \:=  (\pa_i \u{f} \cdot \pa_j \u{f})  
  \es    h_{im} {h_j}^m                                \m .  
\ee 
This is solved by 
\be 
\pa_i f_m = h_{im} \m \mbox{ second-order nilpotent}   \m . 
\ee 
\n Secondly, constant Euclidean derivative yields the PDE  
\be 
c_{ij}   =   D^{Eucl}_{ij}(\u{f})  
   \:=  (\pa_i \u{f} \cdot \pa_j \u{f})  
   \es   h_{im} {h_j}^m                                \m .   
\ee 
A subcase of this 
\be 
( \m c_{ij} = k \delta_{ij} \m )
\ee 
is solved by similar matrices 
\be 
\pa_i f_m =  h_{im} = k \, A_{im}
\ee 
for $k$ a constant scale factor and $A_{im}$ an orthogonal matrix.

\subsection{Equiareal relational derivatives}

Firstly, zero equiareal derivative gives the PDE  
\be 
0  \es  D^{Equi(2)}_{ab} (\u{f}) 
   \es  (  \pa_a \u{f} \cr \pa_b \u{f} )_{\perp}                    \m . 
\ee 
Secondly, constant equiareal derivative yields the PDE  
\be 
c_{ab} \es  D^{Equi(2)}_{ab} (\u{f})  
       \es  ( \pa_a \u{f} \cr \pa_b \u{f} )_{\perp}                 \m . 
\ee

\subsection{Equivoluminal relational derivatives}

Next, zero equivoluminal derivative gives the PDE  
\be 
0  \es  D^{Equi(3)}_{abc} (\u{f})  
   \es  (  \pa_a \u{f} , \,  \pa_b \u{f} ,  \, \pa_c \u{f} )        \m ,  
\ee 
and constant equivoluminal derivative yields the PDE  
\be 
c_{abc} \es  D^{Equi(3)}_{abc} (\u{f})  
        \es  (  \pa_a \u{f} , \,  \pa_b \u{f} ,  \, \pa_c \u{f} )   \m . 
\ee

\subsection{Equi-top-form relational derivatives}

Finally, zero equi-top-form derivative gives the PDE  
\be 
0  \es  D^{Equi(d)}_{a_1 \m . \m . \m . \m  a_d} (\u{f})  
   \es  \bigwedge_{i = 1}^d  \pa_{a_i} \u{f}                                                 \m , 
\ee 
whereas constant equi-top-form derivative yields the PDE  
\be 
c_{a_1 \m . \m . \m . \m  a_d}  \es  D^{Equi(d)}_{a_1 \m . \m . \m . \m  a_d} (\u{f})  
                                \es  \bigwedge_{i = 1}^d  \pa_{a_i} \u{f}                    \m . 
\ee

\section{Conclusion}

\n We succeeded in formulating  relational derivatives -- whether shape derivatives or shape-and-scale derivatives -- in 1-$d$, 
by the simple means of Taylor-expanding the corresponding invariants and finding relationally-invariant factors in the nontrivial leading terms. 
These notions of derivative reflect the underlying minimal nontrivially relational unit (MNRU) supported by each theory. 
The derivative being relationally invariant reflects that it is defined independently of (nondegenerate) choice of the MNRU shape.   

\m 

\n In this way, we found that the 1-$d$ difference derivative is just the ordinary derivative and the 1-$d$ ratio derivative is the logarithmic derivative; both are first-order.  
The 1-$d$ dilatational shape derivative is slightly more involved: the logarithmic derivative of the ordinary derivative, 
as a `direct superposition' of difference and ratio conditions in the form of a second-order derivative.  
Finally, the 1-$d$ projective shape derivative returns the third-order Schwarzian derivative. 
This result firstly confirms that the Schwarzian is defined on an arbitrary (nondegenerate) cluster of 4 points on the line. 
Secondly, it furthermore interprets such a cluster as the MNRU for 1-$d$ projective geometry, 
by which the Schwarzian derivative is confirmed to be a shape-theoretic construct.  

\m 

\n We next posed ODEs for each of these relational derivatives to be zero.  
This returns, respectively, constant functions, nonzero constant functions, nonzero-gradient linear functions and fractional-linear functions. 
The last two of these can be geometrically interpreted as functions lacking ordinary and projective curvature respectively. 
We also posed ODEs for each of these relational derivatives to be constant.  
These remain tractable, returning respectively the linear functions, exponential functions, exponential functions plus constant, and constant plus the tan of a linear function. 

\m 

\n We subsequently considered $\geq$ 2 so as to be entertaining not only clustering information but relative-angle information as well.  
Our simple method succeeds in formulating dilational derivatives and shape-and-scale derivatives: 
translational derivative, dilational derivative, rotational derivative, rotation-and-dilation derivative, Euclidean derivative, and equi-top-form derivative.   

\m 

\n We finally posed PDEs for each of these relational derivatives to be zero.  
The first five of these are solved by, respectively, vector constants (twice), 
                                                     vector functions of constant norm (twice), 
												 and vector functions whose gradients are second-order nilpotent.  
We also posed PDEs for each of these relational derivatives to be constant.  
This returns linear vector functions, 
             an over-determined homogeneous-linear system, 
			 and vector functions of linear squared-norm, 
							         with linear exponential norm, 
                                 and such that their gradient is a similar matrix, respectively.   

\m 

\n{\bf Further research directions}

\m 

\n 0) Formulate $\geq 2$-$d$ dilatational, similarity, conformal, affine and projective shape derivatives. 

\m 

\n Some of the below may be useful in this regard, as well as for gaining further insight into the 1-$d$ relational and $\geq 2$-$d$ 
shape-and-scale derivatives already formulated in the current Article. 

\m 

\n 1) Taking derivatives and Taylor expanding is, more geometrically, prolongation \cite{Cartan55} and consideration of jet bundles \cite{Saunders}. 
The current article considers a shape-theoretic approach to finding differential invariants. 
Its use of a simple expansion method exhibits limitations; we know from elsewhere that e.g. the affine case that our analysis does not reach has 
an affine curvature third-order invariant.
 
\m 

\n 2) Compare with Cartan's \cite{Cartan55} more well-known differential theory of invariants \cite{G63, 9-Pillars}, 
in particular as regards the extent to which this can be interpreted in subsequently-introduced shape-theoretic terms \cite{Kendall84, Small, Kendall, PE16, Minimal-N}.  

\m 

\n{\bf Acknowledgments} I thank Chris Isham and Don Page for previous discussions.  
Reza Tavakol, Malcolm MacCallum, Enrique Alvarez and Jeremy Butterfield for support with my career.  

\begin{appendices}

\section{Lattice of geometrically significant subgroups}
%
{            \begin{figure}[!ht]
\centering
\includegraphics[width=1.0\textwidth]{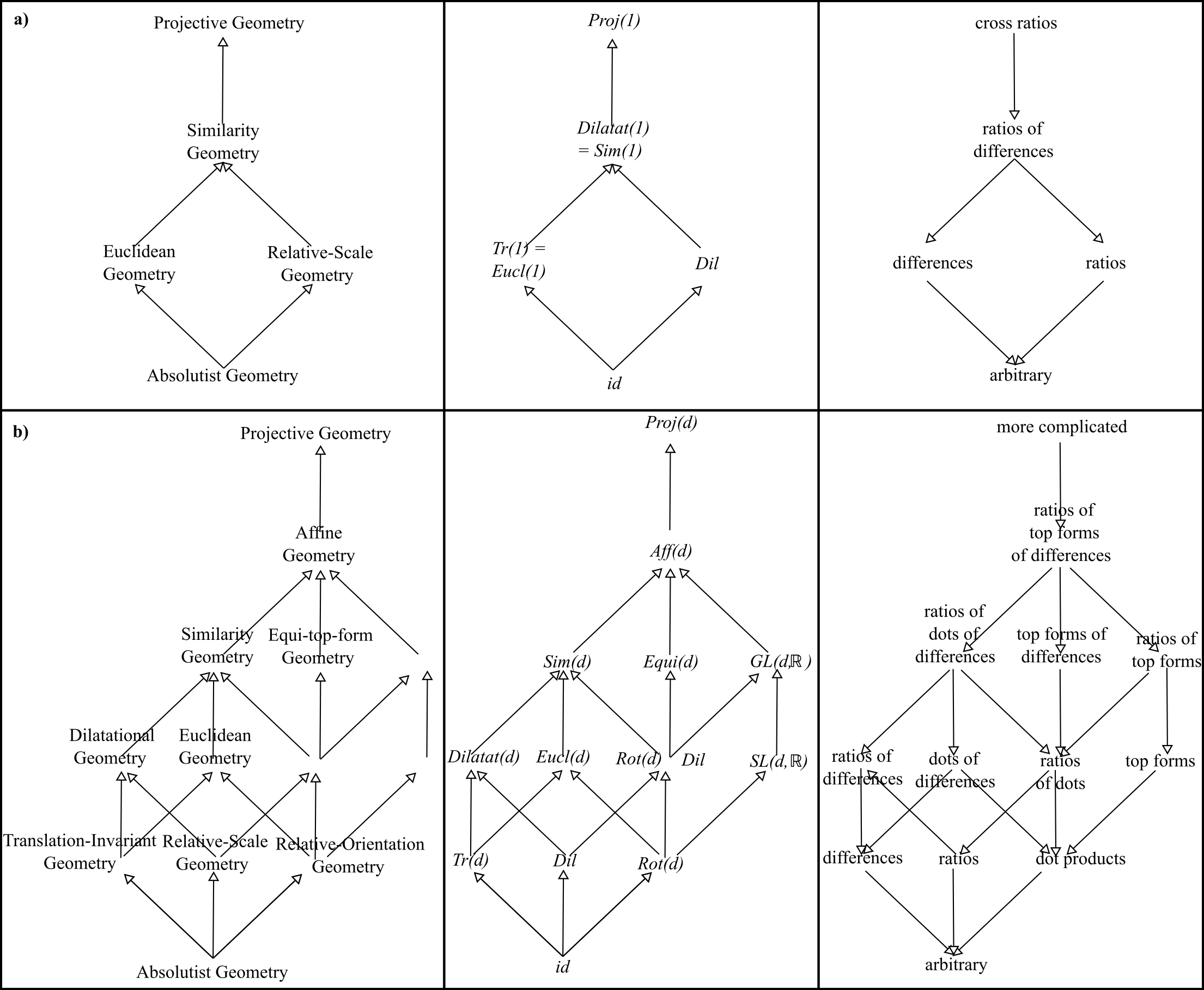}
\caption[Text der im Bilderverzeichnis auftaucht]{        \footnotesize{Bounded lattices of notions of geometry, 
                                                                                their corresponding automorphism groups, 
																			    and the corresponding dual lattice of $N$-point invariants.
Row a) is for 1-$d$ geometries, and row b) is for higher-$d$ geometries whose automorphism group is a subgroup of the projective group.}}
\label{Proj-1-Latt} \end{figure}          }

\end{appendices}

\vspace{10in}


\end{document}